\documentclass[aps,twocolumn,preprintnumbers,amsmath,nofootinbib,amssymb]{revtex4}

\usepackage{graphicx}
\usepackage{times}
\usepackage{float}
\usepackage{color}
\usepackage{amsfonts,amssymb,stmaryrd,latexsym,amsmath}
\usepackage{multirow}
\usepackage{array}
\usepackage{hhline}
\usepackage{booktabs}
\usepackage{enumerate}
\usepackage{slashed}
\usepackage{subfig}
\usepackage{hyperref}
\usepackage{epstopdf}

\makeatletter
\DeclareCaptionJustification{flushboth}{\leftskip\z@skip \rightskip\z@skip \@rightskip\z@skip \parfillskip\@flushglue}
\makeatother
\begin{document}

\hypersetup{
	colorlinks=true,
	citecolor=blue,
	anchorcolor=red,
	menucolor=red,
	linkcolor=red,
	filecolor=red,
	runcolor=red,
	urlcolor=blue,
	frenchlinks=true
}

\title{Novel method for measuring the isovector partner of X(3872)}

\author{Qiu-Lin Tan$^{1}$}
\author{Mao-Jun Yan$^{1}$}\email{yanmj0789@swu.edu.cn}
\author{Chun-Sheng An$^{1}$}
%\thanks{ancs@swu.edu.cn}
\author{Cheng-Rong Deng$^{1}$}
%\thanks{crdeng@swu.edu.cn} 
\author{Xiao-Hai Liu$^{2}$}\email{xiaohai.liu@tju.edu.cn}

\affiliation{$^{1}$School of Physical Science and Technology, Southwest University, Chongqing 400715, China\\
$^{2}$Center for Joint Quantum Studies and Department of Physics, School of Science, Tianjin University, Tianjin 300350, China }

\date{\today}
\begin{abstract}
	The predicted isovector partner of the $X(3872)$, denoted as $W_{c1}$, is expected to produce only a relatively weak structure near the charged $D\bar D^*$ threshold, making its experimental identification challenging. In this work, we explore the triangle singularity (TS) mechanism as a means of enhancing the sensitivity to the $W_{c1}$ in the decays $B^0\to\pi K\bar D^0D^{*0}$. We investigate the $D_{s1}(2536)\bar DD^*$ triangle rescattering mechanism, incorporating the coupled-channel $D\bar D^*$ scattering amplitudes obtained from the Lippmann–Schwinger equation. The interplay between the TS and the coupled-channel dynamics produces characteristic structures in the $K\bar D^0D^{*0}$ invariant mass distributions associated with the $X(3872)$ and its isovector partner $W_{c1}$. In particular, the charged-channel contribution exhibits a distinct structure sensitive to the nearby $W_{c1}$ pole. Our results demonstrate that the TS mechanism provides a promising way to probe the otherwise weak manifestation of the $W_{c1}$, offering a complementary experimental strategy for searching for this predicted state at Belle II and LHCb.
\end{abstract}

\maketitle

\section{Introduction}

In hadron spectroscopy, some peak structures in invariant mass distributions originate from dynamical poles. Kinematic singularities, such as the two-body threshold cusp and triangle singularity (TS), can also produce observations. The former arises from a square-root branch point at an $S$-wave threshold~\cite{Wigner:1948zz}, while the latter develops when three intermediate particles in a triangle diagram can simultaneously go on shell, leading to a logarithmic singularity~\cite{Landau:1959fi}. Under the Coleman-Norton conditions~\cite{Coleman:1965xm}, a TS can approach the physical boundary and generate pronounced structures in invariant mass distributions. Unlike dynamical poles, its location is governed by kinematics. Although such structures may mimic genuine resonances, the strong kinematic sensitivity of a TS can also be exploited to probe nearby dynamical singularities~\cite{Guo:2019twa}, potentially making otherwise weak signals more accessible experimentally.

In recent years, numerous exotic hadrons have been discovered near two-hadron thresholds, suggesting that hadronic molecular components may play an important role in their structure (see the reviews~\cite{Chen:2016qju,Esposito:2016noz,Lebed:2016hpi,Olsen:2017bmm,Ali:2017jda,Guo:2017jvc,Kalashnikova:2018vkv,Liu:2019zoy,Brambilla:2019esw, Bai:2026atm}). Among them, the $X(3872)$, discovered in $B$ decays~\cite{Belle:2003nnu} with quantum numbers $J^{PC}=1^{++}$~\cite{LHCb:2013kgk,LHCb:2015jfc}, is one of the most prominent hadronic molecular candidates. Its mass lies remarkably close to the $\bar D^0D^{*0}$ threshold at approximately $3871.69$~MeV. This proximity, together with its strong coupling to the $D\bar D^*$ channels, suggests that its properties are strongly influenced by near-threshold dynamics. In particular, the mass splitting between the neutral and charged $D\bar D^*$ thresholds makes coupled-channel interactions and isospin-breaking effects essential for understanding the $X(3872)$ and its possible molecular partners.

An intriguing consequence of the coupled-channel $D\bar D^*$ dynamics is the predicted existence of an isovector partner of the $X(3872)$, denoted as $W_{c1}$, with quantum numbers $I^G(J^{PC})=1^-(1^{++})$~\cite{Zhang:2024fxy}. Recently, the connection between the isospin breaking of the $X(3872)$ and the existence of the $W_{c1}$ has been further clarified~\cite{Dong:2026qbr,Shi:2026hwn}. The neutral--charged $D\bar D^*$ threshold splitting provides the seed of isospin breaking, while its substantial amplification requires a sizable interaction in the isovector channel, which also generates a nearby companion pole. However, unlike the pronounced $X(3872)$ signal, the neutral $W_{c1}^0$ is predicted to produce only a relatively weak cusp-like structure near the charged $D\bar D^*$ threshold~\cite{Zhang:2024fxy}. This makes its experimental identification challenging and motivates the search for production mechanisms that can enhance the sensitivity to the predicted $W_{c1}$ pole.

 Triangle singularities offer a promising possibility in this regard. A method for precisely measuring the mass of the $X(3872)$ through the $\gamma X(3872)$ invariant mass distribution was proposed in Ref.~\cite{Guo:2019qcn}. The sensitivity arises from the $D^*\bar D^*D$ triangle loop, where the narrow widths of the intermediate $D^*$ mesons allow the TS to approach the physical boundary, making lineshape highly sensitive to the $X(3872)$ mass. Related mechanisms have been investigated in $\gamma X(3872)$ and $\pi X(3872)$ production~\cite{Sakai:2020crh,Braaten:2019gfj,Braaten:2019gwc,Braaten:2019yua,Sakai:2020ucu}, as well as in the production of the double-charm tetraquark candidate $T_{cc}(3875)$~\cite{Braaten:2022elw}. These studies demonstrate the usefulness of TS mechanisms for probing near-threshold states, particularly through their sensitivity to the masses of the produced hadrons.

In Ref.~\cite{Yan:2022eiy}, a TS mechanism involving the $D_{s1}(2536)\bar DD^*$ triangle loop in $B\to\pi KX(3872)$ decays was proposed to determine the mass of the $X(3872)$, with the $D\bar D^*\to X(3872)$ transition described by a constant coupling $g_X$. While this approach exploits the kinematic sensitivity of the TS to the $X(3872)$ mass, it does not explicitly incorporate the energy dependence of the coupled-channel $D\bar D^*$ scattering amplitude, particularly the charged-threshold dynamics associated with the predicted $W_{c1}$ pole. In the present work, we replace the constant coupling with a two-channel $T$ matrix obtained by solving the Lippmann--Schwinger equation with contact interactions. This simplified framework generates both the $X(3872)$ and $W_{c1}$ poles and allows us to investigate how their associated near-threshold dynamics modify the TS-induced invariant-mass distributions. By varying the isovector contact interaction and examining finite-width effects, we explore the potential of this mechanism to enhance the visibility of the predicted $W_{c1}$ in experimental searches.

In this work, we investigate the decays $B^0\to\pi^-K^+\bar D^0D^{*0}$ and $B^0\to\pi^0K^0\bar D^0D^{*0}$ through the $D_{s1}(2536)\bar DD^*$ triangle rescattering mechanism. We replace the constant coupling $g_X$ used in Ref.~\cite{Yan:2022eiy} with the coupled-channel $T$-matrix obtained from the Lippmann-Schwinger equation. The neutral-channel elastic amplitude $T_{11}$ and the charged-to-neutral transition amplitude $T_{21}$ enter the two decay processes, respectively, allowing the near-threshold dynamics associated with the $X(3872)$ and its predicted isovector partner $W_{c1}$ to be incorporated into the triangle amplitudes. We find distinct structures in the $K\bar D^0D^{*0}$ invariant mass distributions, with the charged-channel contribution exhibiting a characteristic structure sensitive to the nearby $W_{c1}$ pole. These results suggest that the interplay between the TS and coupled-channel dynamics provides a promising approach for experimentally searching for the $W_{c1}$ at Belle II and LHCb.

This paper is organized as follows. In Sec.~\ref{SecII}, we introduce the triangle rescattering mechanism and the coupled-channel $T$-matrix formalism. In Sec.~\ref{sec:results}, we present the numerical results and discuss the predicted invariant mass distributions. A summary and discussion are given in Sec.~\ref{sec:summary}.

\section{the triangle diagram and coupled channel scattering}\label{SecII}

\begin{figure}[htb]
 \centering
 \includegraphics[width=0.45\textwidth]{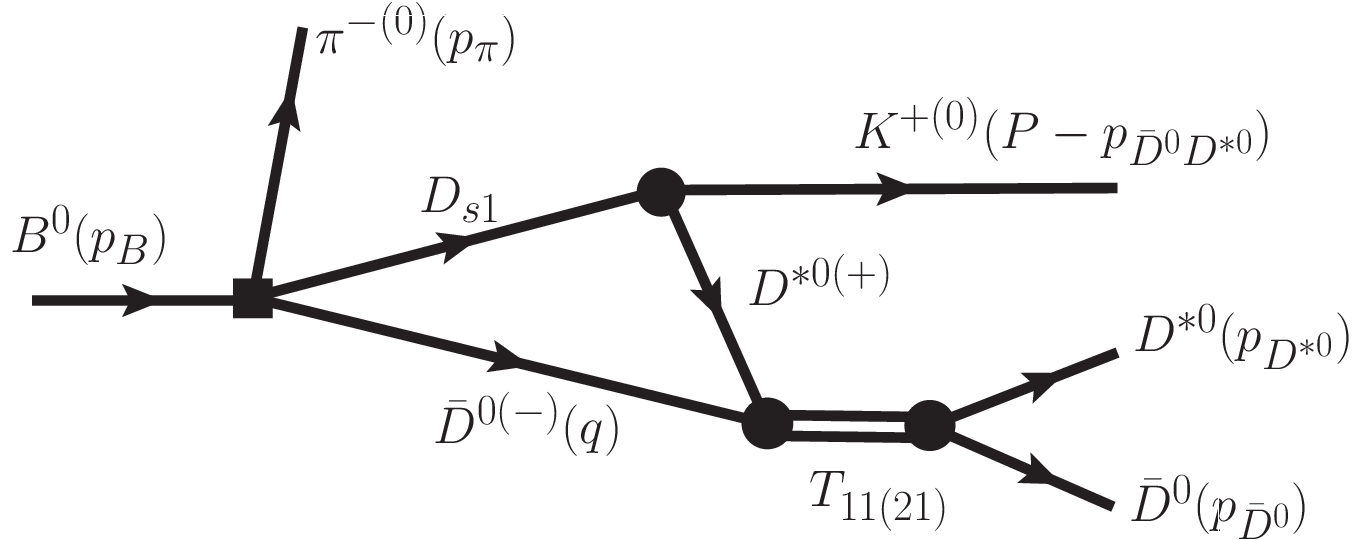}
 \caption{Triangle rescattering mechanisms for
 $B^0\to\pi^-K^+\bar D^0D^{*0}$ (neutral intermediate channel) and
 $B^0\to\pi^0K^0\bar D^0D^{*0}$ (charged intermediate channel).
 In each case $P=p_B-p_\pi=p_K+p_{\bar D^0}+p_{D^{*0}}$ is the
 four-momentum of the $K\bar D^0D^{*0}$ system.}
 \label{fig:triangle}
\end{figure}

We investigate the decays $B^0\to\pi^-K^+\bar D^0D^{*0}$ and
$B^0\to\pi^0K^0\bar D^0D^{*0}$ through the
$D_{s1}(2536)\bar DD^*$ triangle rescattering mechanism, as shown in
Fig.~\ref{fig:triangle}. The first process involves the intermediate
$D_{s1}^+\bar D^0D^{*0}$ state followed by neutral-channel elastic
rescattering, whereas the second proceeds through
$D_{s1}^+D^-D^{*+}$ and the charged-to-neutral transition
$D^-D^{*+}\to\bar D^0D^{*0}$. The kinematic conditions for the triangle
singularity (TS) in this mechanism have been analyzed in
Ref.~\cite{Yan:2022eiy}. For reference, the corresponding TS regions are
\begin{align}
 m_{\bar D^0D^{*0}}&\in[3871.69,3875.24]~\mathrm{MeV}, \nonumber\\
 M_{K^+\bar D^0D^{*0}}&\in[4399.93,4403.66]~\mathrm{MeV},\label{eq:TSneutral}\\
 m_{D^-D^{*+}}&\in[3879.92,3882.85]~\mathrm{MeV}, \nonumber\\
 M_{K^0D^-D^{*+}}&\in[4404.77,4407.71]~\mathrm{MeV}.\label{eq:TScharged}
\end{align}
The two regions probe different parts of the $D\bar D^*$ coupled-channel
amplitude. In particular, the charged-channel triangle provides access
to the dynamics near the $D^-D^{*+}$ threshold, where the predicted
$W_{c1}$ may affect the rescattering amplitude.

To incorporate the near-threshold dynamics associated with the $X(3872)$ and its predicted isovector partner, we replace the constant coupling $g_X$ used in Ref.~\cite{Yan:2022eiy} with a two-channel $T$ matrix constructed from contact interactions by solving the Lippmann--Schwinger equation. The contact couplings are chosen to reproduce the pole positions reported in Ref.~\cite{Ji:2025hjw}. This approximation retains the energy dependence and coupled-channel threshold structure relevant to the present study, while not incorporating the full dynamical ingredients of Ref.~\cite{Ji:2025hjw}. Denoting the neutral $\bar D^0D^{*0}$ and
charged $D^-D^{*+}$ channels by indices 1 and 2, respectively, the
on-shell amplitude is written as
\begin{eqnarray}
 T(E)&=&\bigl[1-VG(E)\bigr]^{-1}V,\label{eq:Tmatrix}\\
V&=&\frac12\begin{pmatrix}
 C_{0X}+C_{1X}&C_{0X}-C_{1X}\\
 C_{0X}-C_{1X}&C_{0X}+C_{1X}
 \end{pmatrix},
 \label{eq:Tcompact}
\end{eqnarray}
where $G(E)=\operatorname{diag}(G_1(E),G_2(E))$ denotes the regularized
nonrelativistic two-body loop functions, and \(C_{0X}=-11.11^{+0.15}_{-0.15}\,\mathrm{GeV}^{-2}\)
and
\(C_{1X}=-8.67^{+0.10}_{-0.10}\,\mathrm{GeV}^{-2}\)
are the isoscalar and isovector contact couplings, respectively, which are regularized
with
$q_{\mathrm{max}}=1~\mathrm{GeV}$ by fitting to the poles
in Ref.~\cite{Ji:2025hjw}. Here $E$ is the energy relative to
the neutral threshold; $T_{11}(E)$ represents neutral-channel elastic
scattering and $T_{21}(E)$ the charged-to-neutral transition. Their
energy dependence, rather than a constant coupling, is essential for
retaining the charged-threshold structure in the production amplitude.

We consider the decay amplitudes for the processes $B^0\to\pi^-K^+\bar D^0D^{*0}$ and $B^0\to\pi^0K^0\bar D^0D^{*0}$, which receive contributions from the two triangle rescattering diagrams shown in Fig.~\ref{fig:triangle}. Although the final states have the same hadronic content, the intermediate $D\bar D^*$ channels differ: the first process involves elastic rescattering in the neutral channel, $\bar D^0D^{*0}\to\bar D^0D^{*0}$, whereas the second involves the transition from the charged channel, $D^-D^{*+}\to\bar D^0D^{*0}$.

The relevant vertex factors are specified as follows. The production vertex is parametrized as
\begin{eqnarray}
t_{B^0\to\pi D_{s1}\bar D^0}
&=&
\mathcal C_1\,
\vec\epsilon^{\,*}(D_{s1})\cdot\vec p_\pi,
\label{eq:vertexB}
\end{eqnarray}
where $\vec\epsilon(D_{s1})$ denotes the polarization vector of the $D_{s1}(2536)$ and $\vec p_\pi$ is the pion momentum. In the $D_{s1}\bar D^0$ center-of-mass frame, the total three-momentum of the pair vanishes, so that $\vec p_B-\vec p_\pi=0$. The coupling $\mathcal C_1$ is estimated from the relevant experimental branching fractions; its numerical value is taken to be approximately $7.0\times10^{-6}$ in the present calculation.

The $D_{s1}(2536)\to D^*K$ vertex is given by
\begin{eqnarray}
t_{D_{s1}\to D^*K}
&=&
g_{D_{s1}D^*K}\,
\vec\epsilon(D_{s1})\cdot
\vec\epsilon^{\,*}(D^*),
\label{eq:vertexDs1}
\end{eqnarray}
where $g_{D_{s1}D^*K}\simeq0.78~\mathrm{GeV}$ is determined from experimental data~\cite{ParticleDataGroup:2026mpi}. The $D_{s1}(2536)$ is a narrow state, with $\Gamma_{D_{s1}}=0.92\pm0.05~\mathrm{MeV}$~\cite{ParticleDataGroup:2026mpi}, making its width potentially important for the TS-induced lineshape. The broader $D_{s2}(2573)$ contribution is not included in this work.

The  $D\bar D^*$ rescattering amplitudes are parametrized as
\begin{eqnarray}
t_{\bar D^0D^{*0}\to\bar D^0D^{*0}}
&=&
T_{11}\,
\vec\epsilon(D^{*0}_{\rm in})\cdot
\vec\epsilon^{\,*}(D^{*0}_{\rm out}),
\nonumber\\
t_{D^-D^{*+}\to\bar D^0D^{*0}}
&=&
T_{21}\,
\vec\epsilon(D^{*+}_{\rm in})\cdot
\vec\epsilon^{\,*}(D^{*0}_{\rm out}),
\label{eq:vertexT}
\end{eqnarray}
with the polarization vector $\vec{\epsilon}(D^{\ast}_{\rm in/out})$ of the incoming/outgoing $D^{\ast}$,
where $T_{11}$ and $T_{21}$ are the coupled-channel $T$-matrix elements in Eq. (\ref{eq:Tmatrix}). %They replace the constant effective coupling $g_X$ used in Ref.~\cite{Yan:2022eiy}, thereby retaining the energy dependence associated with the $D\bar D^*$ final-state interaction.

The  vertexes are given above. Summing over the polarizations of the internal $D_{s1}$ and $D^*$ mesons, and using the nonrelativistic approximation $\sum_\lambda \epsilon_i(\lambda)\epsilon_j^*(\lambda)\simeq\delta_{ij}$, the internal polarization structures contract to leave only the polarization vector of the external $D^{*0}$. Since the adopted $D_{s1}D^*K$ vertex contains no explicit momentum dependence, the remaining loop integral is scalar. The two decay amplitudes can therefore be expressed in a unified form,
\begin{equation}
\mathcal M_i(s,t)=
\mathcal C_i g_i\,T_{i1}
\bigl[\vec p_{\pi_i}\cdot
\vec\epsilon^{\,*}(D^{*0}_{\rm out})\bigr]
I_\triangle^{(i)}(s,t),
\,\, i=1,2,
\label{eq:amplitudecompact}
\end{equation}
where $\mathcal C_i$ and $g_i$ denote the production and $D_{s1}D_i^*K_i$ couplings, respectively, for each charge channel. The amplitudes $\mathcal M_1$ and $\mathcal M_2$ correspond to the neutral- and charged-channel triangle mechanisms, with $T_{11}$ and $T_{21}$ describing the corresponding $D\bar D^*$ rescattering processes.

The scalar triangle integral is defined as
\begin{eqnarray}
I_\triangle^{(i)}(s,t)
&=&i\!\int\!\frac{d^4q}{(2\pi)^4}
\frac{1}{
[(P-q)^2-m_{D_{s1}}^2]
[q^2-m_{D_i}^2]
}
\nonumber\\
&&\times
\frac{1}{
[(p_{\bar D^0D^{*0}}-q)^2-m_{D_i^*}^2]
},
\label{eq:trianglecompact}
\end{eqnarray}
where $(D_1,D_1^*)=(\bar D^0,D^{*0})$ and
$(D_2,D_2^*)=(D^-,D^{*+})$. The kinematic variables are
$s=M_{K\bar D^0D^{*0}}^2$,
$t=M_{\bar D^0D^{*0}}^2$, and
$E=\sqrt{t}-m_{\bar D^0}-m_{D^{*0}}$.
Thus, the two mechanisms differ in both the rescattering amplitude and the intermediate-particle masses entering the triangle loop. Following Ref.~\cite{Yan:2022eiy}, we evaluate the loop integral using \textsc{LoopTools}~\cite{Hahn:2000jm}.

The observable distributions are obtained from
\begin{equation}
 \frac{d^2\Gamma_i}{dM_{K\bar D^0D^{*0}}\,dM_{\bar D^0D^{*0}}}
 =\frac{1}{(2\pi)^5}\frac{\widetilde p_i\,p_{K_i}^{*}\,p_D^{**}}
 {4m_B^2}\,\overline{|\mathcal M_i|^2},\, i=1,2,
 \label{eq:widthcompact}
\end{equation}
where $\overline{|\mathcal M_i|^2}$ includes the polarization sum and
\begin{align}
 \widetilde p_i&=\frac{\lambda^{1/2}(m_B^2,m_{\pi_i}^2,s)}{2m_B},\\
 p_{K_i}^{*}&=\frac{\lambda^{1/2}(s,m_{K_i}^2,t)}{2\sqrt s},\\
 p_D^{**}&=\frac{\lambda^{1/2}(t,m_{\bar D^0}^2,m_{D^{*0}}^2)}{2\sqrt t},
\end{align}
with $\lambda(x,y,z)=x^2+y^2+z^2-2xy-2yz-2zx$. Integrating
Eq.~\eqref{eq:widthcompact} over
\begin{equation}
 m_{\bar D^0}+m_{D^{*0}}\le M_{\bar D^0D^{*0}}
 \le M_{K\bar D^0D^{*0}}-m_{K_i}
\end{equation}
yields $d\Gamma_i/dM_{K\bar D^0D^{*0}}$, which we use to study
how the TS mechanism makes the charged-channel dynamics associated
with the predicted $W_{c1}$ accessible in the two decay modes.

\section{Numerical results and discussion}
\label{sec:results}

We now examine the $K\bar D^0D^{*0}$ invariant-mass distributions in the two
$B^0$ decay modes introduced in Sec.~\ref{SecII}. The purpose is not to
redetermine the $X(3872)$ or $W_{c1}$ pole positions, but to investigate
whether the $D_{s1}(2536)\bar DD^*$ triangle mechanism provides an
experimentally useful probe of the charged-channel dynamics associated with
the predicted $W_{c1}$. We use the contact-interaction approximation to the
coupled-channel amplitude, with $q_{\rm max}=1~\mathrm{GeV}$ and the
low-energy constants adopted from Ref.~\cite{Ji:2025hjw}.
Accordingly, the lineshapes presented below are predictions conditional on
this particular scattering input, rather than an independent determination of
the existence of the $W_{c1}$.

For each process, the $\bar D^0D^{*0}$ invariant mass is integrated over
its kinematically allowed range to obtain
$d\Gamma_i/dM_{K\bar D^0D^{*0}}$. The neutral intermediate state contributes
to $B^0\to\pi^-K^+\bar D^0D^{*0}$ through $T_{11}$, whereas the charged
intermediate state contributes to $B^0\to\pi^0K^0\bar D^0D^{*0}$ through
$T_{21}$. In both cases, the production amplitude contains the product of a
triangle loop and the energy-dependent rescattering amplitude. In particular,
$\mathcal M_2\propto I_\triangle^{(2)}T_{21}$, so the charged-channel
threshold dependence is retained rather than replaced by a constant
$D\bar D^*$ transition vertex.

%Figures display the corresponding invariant-mass distributions. The neutral-channel mechanism produces a relatively sharp structure, while the charged-channel mechanism gives a broader enhancementat a higher $K\bar D^0D^{*0}$ invariant mass. The shift reflects the different intermediate-particle masses and the associated TS kinematics; the detailed widths and shapes also depend on the respective $T$-matrix elements and the integration over the $D\bar D^*$ invariant mass. The charged-channel structure is of particular interest because $T_{21}$ is sensitive to the region of the charged $D\bar D^*$ threshold, where an isovector companion pole has been predicted. It should nevertheless be stressed that a TS can generate a kinematic enhancement even without a nearby $W_{c1}$ pole. Thus, the mere appearance of the second structure is not, by itself, evidence for a new state: the information on the $W_{c1}$ is encoded in the dependence of its lineshape on the charged-channel scattering amplitude. 

We firstly investigate the finite-width effects of the intermediate
$D^*$ and $D_{s1}(2536)$ mesons in the triangle diagrams. These effects
are relevant for the $K\bar D^0D^{*0}$ invariant-mass
distribution, whose lineshape can be sensitive to the mass and width
of the $D_{s1}(2536)$~\cite{Wang:2026yyx}. We adopt two prescriptions
to account for the $D_{s1}(2536)$ width, while describing the $D^*$
mass distribution using a Breit--Wigner form in both cases. In Case I,
the $D_{s1}(2536)$ mass is varied between
$m_{D_{s1}}^{\rm phy.}-\Gamma_{D_{s1}}/2$ and
$m_{D_{s1}}^{\rm phy.}+\Gamma_{D_{s1}}/2$, where
$m_{D_{s1}}^{\rm phy.}$ and $\Gamma_{D_{s1}}$ are the PDG values.
In Case II, the $D_{s1}(2536)$ is assigned a complex mass,
$m_{D_{s1}}^{\rm phy.}-i\Gamma_{D_{s1}}/2$, to investigate how its
finite width smears the structures generated by the triangle
mechanism. The complex-mass prescription is used to study
finite-width effects in hadron spectroscopy
\cite{Jing:2019cbw,Yan:2022eiy,Yan:2025bez}.

\begin{figure}[htb]
  \centering
  \includegraphics[width=0.45\textwidth]{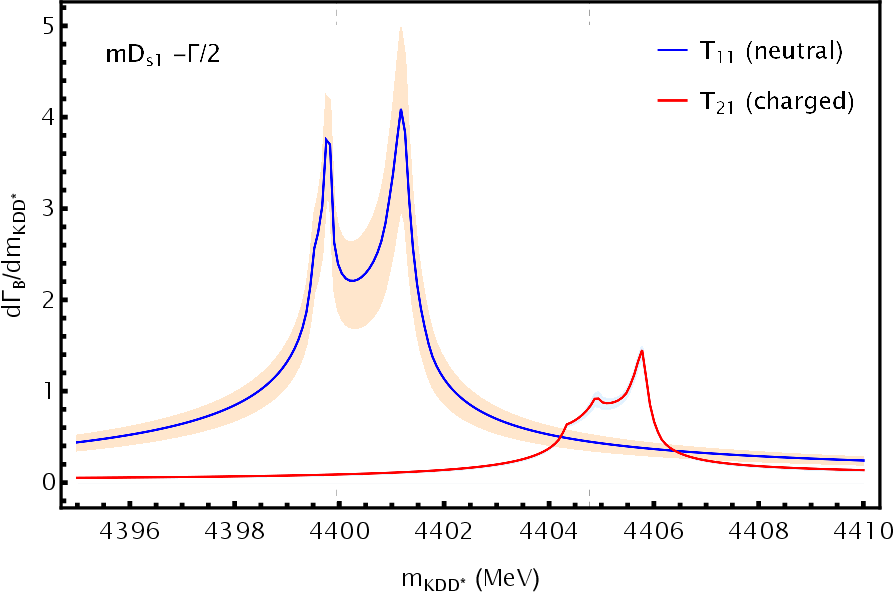}
  \includegraphics[width=0.45\textwidth]{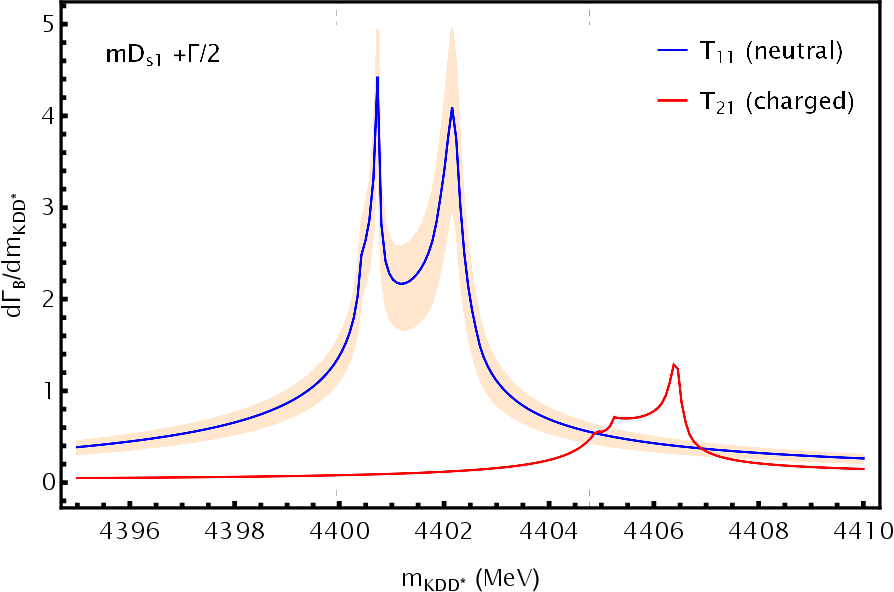}
  \caption{Predicted $M_{K\bar D^0D^{*0}}$ distributions from the
  $D_{s1}(2536)\bar DD^*$ triangle mechanisms, including the finite
  widths of the intermediate particles. The blue and red curves
  correspond to $B^0\to\pi^-K^+\bar D^0D^{*0}$ with neutral-channel
  rescattering ($T_{11}$) and $B^0\to\pi^0K^0\bar D^0D^{*0}$ with
  charged-to-neutral rescattering ($T_{21}$), respectively. The
  relative heights of the two distributions depend on the
  channel-dependent production amplitudes and should not be interpreted
  as a model-independent event-yield ratio.}
  \label{fig:TSDs1Mass}
\end{figure}

For Case I, the results obtained with the two limiting values of $m_{D_{s1}}$ are shown in Fig.~\ref{fig:TSDs1Mass}, where the bands correspond to the uncertaities of $C_{0X}$ and $C_{1X}$. The neutral-channel contribution exhibits two prominent structures. The lower-mass peak is associated with the near-threshold $X(3872)$ pole, while the higher-mass peak is associated with the TS. In the charged-channel contribution, the cusp near the charged $D\bar D^*$ threshold reflects the virtual-state dynamics associated with the $W_{c1}$ pole, whereas the higher-energy enhancement is primarily driven by the TS. The detailed lineshapes result from the interplay between the triangle mechanism and the corresponding coupled-channel amplitudes. Varying $m_{D_{s1}}$ shifts the structures, illustrating the sensitivity of the predicted distributions to the intermediate-particle mass. In particular, the charged-channel lineshape may provide a useful probe of the predicted $W_{c1}$ dynamics.

\begin{figure}[htb]
  \centering
  \includegraphics[width=0.45\textwidth]{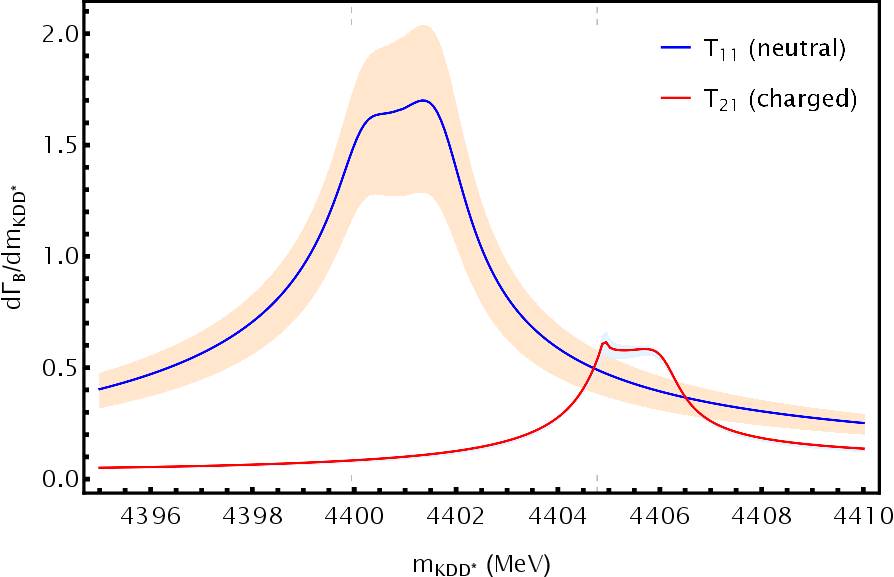}
  
  \caption{Predicted $M_{K\bar D^0D^{*0}}$ distributions from the
  $D_{s1}(2536)\bar DD^*$ triangle mechanisms, including the finite
  widths of the intermediate particles. The blue and red curves
  correspond to $B^0\to\pi^-K^+\bar D^0D^{*0}$ with neutral-channel
  rescattering ($T_{11}$) and $B^0\to\pi^0K^0\bar D^0D^{*0}$ with
  charged-to-neutral rescattering ($T_{21}$), respectively. The
  relative heights of the two distributions depend on the
  channel-dependent production amplitudes and should not be interpreted
  as a model-independent event-yield ratio.}
  \label{fig:TSlineshape}
\end{figure}

For Case II, the complex-mass prescription smears the sharp structures
found in Case I. As shown in Fig.~\ref{fig:TSlineshape}, the nearby
structures associated with the near-threshold dynamics and the TS
merge into a shoulder-like enhancement because the $D_{s1}(2536)$
width is comparable to their separation. Nevertheless, the enhancement
in the charged-to-neutral rescattering contribution remains visible.
Thus, the finite width modifies the detailed lineshape and reduces
the sharpness of the structures, but does not eliminate the
TS-related enhancement in the charged-channel distribution. The blue
and red curves correspond to different $B^0$ decay channels and
exhibit structures in different kinematic regions. Their relative
heights depend on the channel-dependent production amplitudes and
should not be interpreted as a model-independent event-yield ratio.

This result motivates a further investigation of the sensitivity
of the TS-induced lineshapes to the isovector $D\bar D^*$ interaction.
In the following, we vary the isovector contact coupling $C_{1X}$
while keeping $C_{0X}$ fixed, and examine how the resulting motion
of the $W_{c1}$ pole affects the observable invariant-mass
distributions.

\begin{figure}[htb]
  \centering
  \includegraphics[width=0.4\textwidth]{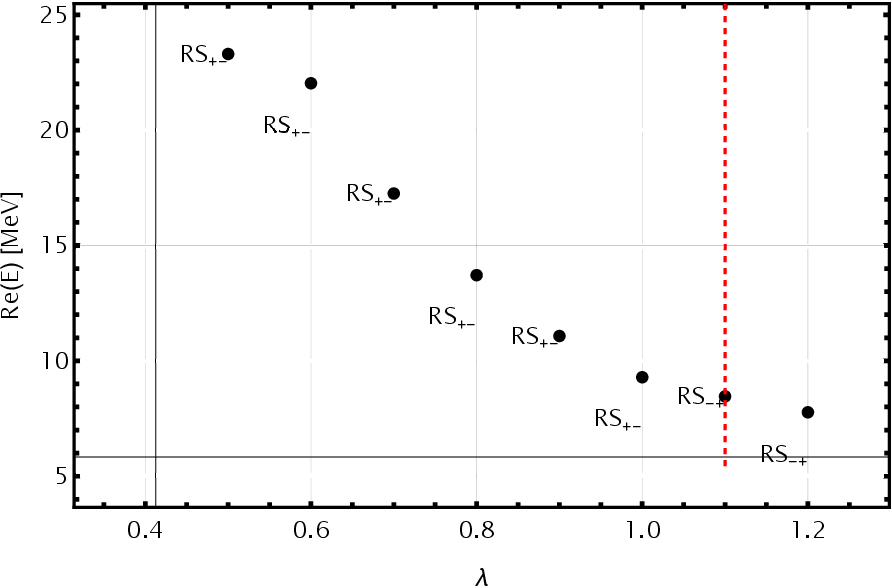}
  \includegraphics[width=0.4\textwidth]{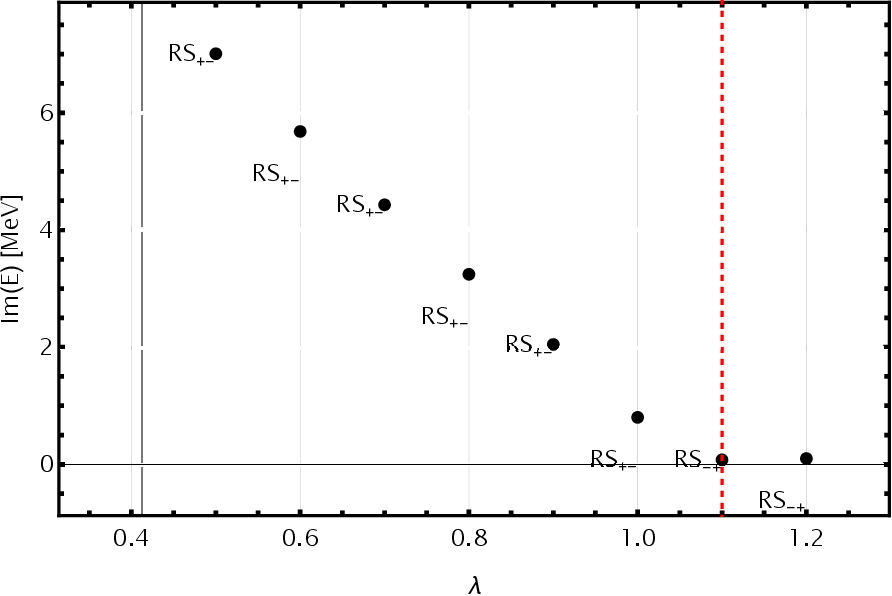}
  \caption{Real and imaginary parts of the $W_{c1}$ pole as
  functions of the scaling parameter $\lambda$, defined by
  $C_{1X}\to\lambda C_{1X}$ with $C_{0X}$ fixed. The $+/-$ stands for the first/second Riemann sheet of the corresponding channel.}
  \label{fig:Wc1Pole}
\end{figure}

It is instructive to compare the present analysis with existing determinations of the isovector $D\bar D^*$ interaction. The dispersive analysis of Ref.~\cite{Dias:2024zfh} constrains the isospin-breaking dynamics of the $X(3872)$ through its $J/\psi\pi^+\pi^-$ decay, extracting the coupling ratio $R_X=g_{X\psi\rho}/g_{X\psi\omega}=0.26\pm0.03$ and updating the predictions for the $W_{c1}$ poles. In particular, it reports a neutral $W_{c1}^0$ pole near $3881.7~\mathrm{MeV}$, approximately $1.8~\mathrm{MeV}$ above the charged $D\bar D^*$ threshold on an unphysical sheet. A more comprehensive coupled-channel analysis was performed in Ref.~\cite{Ji:2025hjw}, which simultaneously fits BESIII and LHCb data and incorporates the energy-dependent $D^*$ self-energy, the associated $D\bar D\pi$ three-body dynamics, one-pion exchange, and inelastic channels. These studies provide important constraints on the isovector interaction, but their dynamical descriptions are more elaborate than the contact-interaction model adopted here.

In the present work, we use an energy-independent contact interaction parametrized by $C_{0X}$ and $C_{1X}$ to reproduce the poles obtained in Ref.~\cite{Ji:2025hjw}. This simplified description does not explicitly include one-pion exchange, the full $D\bar D\pi$ three-body dynamics, or inelastic channels, and therefore cannot capture all the dynamical effects incorporated in the more comprehensive analysis. To explore how such model simplifications may affect the predicted $W_{c1}$ pole and its observable manifestations, we rescale the isovector coupling as $C_{1X}\to\lambda C_{1X}$ while keeping $C_{0X}$ fixed. We then track the resulting pole trajectory and examine its impact on the TS-induced invariant-mass distributions. %This procedure probes the sensitivity of the pole and lineshapes to variations of the isovector interaction within our simplified framework. Although it allows us to explore potentially substantial shifts in the $W_{c1}$ pole, it does not provide a rigorous upper bound on the theoretical uncertainty associated with omitted dynamics or alternative interaction models.

To investigate the $W_{c1}$ pole trajectory, the rescaled isovector contact
coupling is adopted.
The resulting $W_{c1}$ pole positions are shown in
Fig.~\ref{fig:Wc1Pole}. As $\lambda$ increases, the isovector
interaction becomes stronger and the pole approaches the charged
$D\bar D^*$ threshold for $\lambda<1.1$. For larger values of
$\lambda$, the pole moves onto the $RS_{-+}$ sheet, which is physical
with respect to the charged channel.
The pole thus acquires a bound-state-like character relative to the
charged threshold. We restrict the scan to $\lambda\leq1.2$; beyond
this range, the rescaled isovector interaction becomes stronger than
the isoscalar interaction in the parameterization adopted here.

This pole trajectory provides an estimate of the interaction dependence
within the present contact-interaction model. The real part of the pole
relative to the threshold determines its location with respect to the
threshold, while its imaginary part and residue influence how the pole
manifests itself in the physical scattering amplitude. Consequently,
the pole position alone does not uniquely determine the observable
lineshape, and its impact must be examined by calculating the
corresponding production amplitudes.

\begin{figure}[htb]
  \centering
  \includegraphics[width=0.4\textwidth]{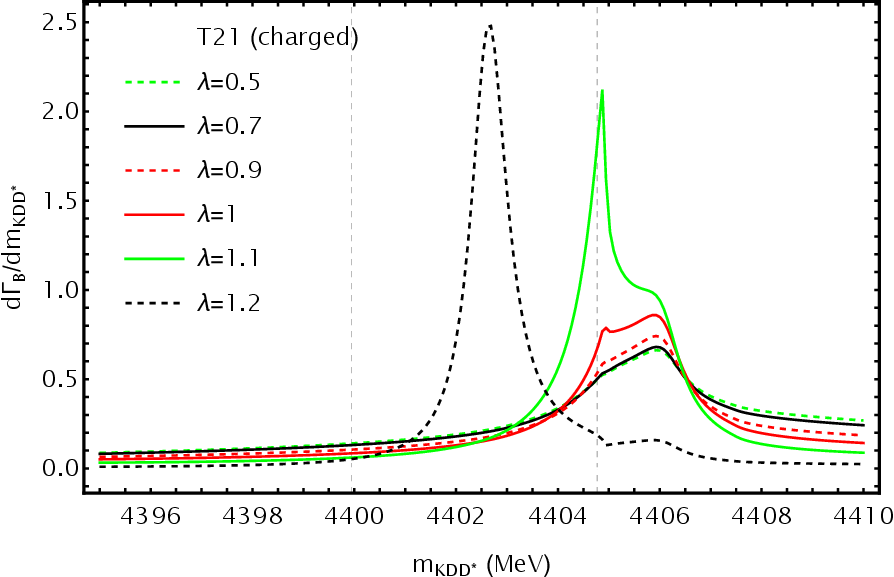}
  \caption{Predicted $M_{K^0\bar D^0D^{*0}}$ distributions for
  representative values of $\lambda$, with
  $C_{1X}\to\lambda C_{1X}$ and $C_{0X}$ fixed. The same finite-width
  treatment is used for all curves. The variation illustrates how
  changes in the isovector interaction and the associated $W_{c1}$
  pole affect the TS-induced invariant-mass distributions.}
  \label{fig:TSvarying}
\end{figure}

We now examine how the variation of $C_{1X}$ affects the observable invariant-mass distributions. Figure~\ref{fig:TSvarying} shows the spectra obtained for representative values of $\lambda$, using the same finite-width treatment. A nontrivial enhancement appears in the $4402$--$4406~\mathrm{MeV}$ region. For $\lambda\leq0.9$, the virtual $W_{c1}$ pole remains relatively far from the charged threshold, and the cusp in $T_{21}$ is comparatively mild, so the resulting lineshapes largely overlap. As $\lambda$ increases, the pole approaches the charged threshold, leading to a more pronounced threshold enhancement around $4405~\mathrm{MeV}$. For $\lambda\gtrsim1.1$, the pole develops a bound-state-like character with respect to the charged channel, and the enhancement becomes more prominent. At $\lambda=1.2$, the dominant enhancement shifts toward $4403~\mathrm{MeV}$. This behavior reflects the interplay between the evolving near-threshold structure of $T_{21}$ and the TS kinematics, indicating that the proposed production mechanism can be sensitive to the qualitative nature of the isovector interaction. The magnitude of this sensitivity, however, depends on the adopted interaction model.

The predicted lineshapes should be regarded as illustrative rather than direct predictions of experimental event yields. The relative yields of the two $B^0$ decay channels depend on their respective weak-production amplitudes and cannot be inferred from the normalized spectra alone. A realistic assessment of the experimental sensitivity to the $W_{c1}$ therefore requires accounting for non-TS backgrounds, interference effects, detector resolution, and event-selection efficiencies.

Beyond the specific case of the $W_{c1}$, the present strategy may be extended to other near-threshold hadronic states whose signals are sensitive to coupled-channel dynamics. By combining suitable triangle-singularity kinematics with energy-dependent scattering amplitudes, one can investigate how nearby poles and threshold effects manifest themselves in experimentally accessible invariant-mass distributions. This approach may be particularly useful for hadronic molecular candidates, for which the interplay between kinematic singularities and near-threshold poles can produce distinctive, channel-dependent lineshapes. The key requirement is to identify production processes in which the triangle kinematics enhance the sensitivity to the relevant coupled-channel dynamics. Such applications would provide a complementary means of probing near-threshold poles and exploring the nature of exotic hadrons, e.g. $P_c(4457)$ \cite{Guo:2019fdo,Nakamura:2021qvy,Peng:2022iez,Yang:2022ezl,Wang:2026yyx} and $P_{cs}(4338)$ \cite{Burns:2022uha,Nakamura:2022gtu,Yan:2022wuz,Yang:2022ezl}.

%In summary, the finite-width study shows that the TS-related enhancement survives the smearing considered here, while the $C_{1X}$ scan demonstrates that the predicted lineshape depends on the nature of the near-threshold $W_{c1}$ pole. In particular, the spectra vary relatively little within the virtual-state regime but change more substantially when the pole develops a bound-state-like character. These findings support the use of the $K\bar D^0D^{*0}$ invariant-mass distribution as a complementary probe of the isovector $D\bar D^*$ interaction. Nevertheless, a quantitative assessment of the experimental sensitivity requires more complete coupled-channel amplitudes, realistic backgrounds, and detector effects. The present results should therefore be viewed as a proof of principle for a complementary $W_{c1}$ search channel, rather than as evidence that the isovector partner has already been observed.

\section{Summary}\label{sec:summary}

In this work, we have investigated the decays $B^0\to\pi^-K^+\bar D^0D^{*0}$ and $B^0\to\pi^0K^0\bar D^0D^{*0}$ via the $D_{s1}(2536)\bar DD^*$ triangle rescattering mechanism. By replacing the constant coupling $g_X$ used in Ref.~\cite{Yan:2022eiy} with an energy-dependent two-channel $T$ matrix constructed from contact interactions, we investigate how the near-threshold dynamics associated with the $X(3872)$ and its predicted isovector partner $W_{c1}$ affect the TS-induced invariant-mass distributions. We find that the TS-related enhancements remain visible under the finite-width prescriptions considered in this work. Furthermore, varying the isovector contact interaction reveals that the lineshapes are relatively stable when the $W_{c1}$ pole remains in the virtual-state regime, but change more substantially when the pole develops a bound-state-like character with respect to the charged channel. These results demonstrate how the interplay between triangle kinematics and coupled-channel dynamics can influence experimentally accessible invariant-mass distributions, and suggest that the proposed mechanism may provide a complementary approach to searching for the $W_{c1}$. A quantitative assessment of its experimental sensitivity will require a more complete treatment of the coupled-channel dynamics, non-TS backgrounds, interference effects, and detector resolution. The present study therefore provides a proof of principle for this search strategy, rather than an independent determination of the $W_{c1}$ pole parameters or evidence for its experimental observation.

\begin{acknowledgments}
This research is supported by the National Natural Science Foundation of China under Grants No. 12305096, No. 12235018 and No. 12675106, the Fundamental Research Funds for the Central Universities under Grant
No. SWU-XDJH202304, No. SWU-KQ25016 and Chongqing Natural Science Foundation under Project No. CSTB2025NSCQ-GPX0516.
\end{acknowledgments}

%\bibliographystyle{apsrev4-1}
%\bibliography{wenxian}

\end{document}